# Clustering-Based Predictive Process Monitoring

Chiara Di Francescomarino, Marlon Dumas, Fabrizio Maria Maggi and Irene Teinemaa
FBK-IRST, Via Sommarive 18,38050 Trento, Italy.
dfmchiara@fbk.eu
University of Tartu, Liivi 2, 50409 Tartu, Estonia
{marlon.dumas, f.m.maggi, teinemaa}@ut.ee

**Abstract**

*Business process enactment is generally supported by information systems that record data about process executions, which can be extracted as event logs. Predictive process monitoring is concerned with exploiting such event logs to predict how running (uncompleted) cases will unfold up to their completion. In this paper, we propose a predictive process monitoring framework for estimating the probability that a given predicate will be fulfilled upon completion of a running case. The predicate can be, for example, a temporal logic constraint or a time constraint, or any predicate that can be evaluated over a completed trace. The framework takes into account both the sequence of events observed in the current trace, as well as data attributes associated to these events. The prediction problem is approached in two phases. First, prefixes of previous traces are clustered according to control flow information. Secondly, a classifier is built for each cluster using event data to discriminate between fulfillments and violations. At runtime, a prediction is made on a running case by mapping it to a cluster and applying the corresponding classifier. The framework has been implemented in the ProM toolset and validated on a log pertaining to the treatment of cancer patients in a large hospital.*

## 1 Introduction

Everyday life routinely poses questions and challenges on prediction: "how much longer is my 10 years old cat expected to live?"; "will the 12:00 noon fast train from Rome to Milan arrive on time so that I can catch my connection?"; "what will the share price of my investment be in 10 days?". For all these questions asking whether in the future a certain event will occur or not (a certain property will be satisfied or not), we cannot expect to provide a true answer based on what we already know. Yet, what we learned from history and past experiences often suggests a reasonable answer.

Analogous questions and predictive challenges can arise during the execution of business processes. For example, in a medical process execution a doctor may ponder whether a surgery, a pharmacological therapy or a manipulation is the best choice to be made in order to guarantee the patient recovery. *Predictive business process monitoring* [12] is a family of techniques that apply what we do in everyday life to the field of business processes. In particular, predictive process monitoring exploits event logs, which are more and more widespread in modern information systems, to predict how current (uncompleted) cases will unfold up to their completion. A predictive process monitor allows users to specify predicates, for example using Linear Temporal Logic (LTL) or any other language, to capture boolean functions over traces of completed cases. Based on the analysis of execution traces, the monitor continuously provides the user with estimations of the likelihood of achieving the given predicates for a running case.

New opportunities to foster and improve the performance of predictive business process monitoring are provided by the fast growing availability of data. In the so-called "big data" era, the enormous quantities of various and heterogeneous data produced and managed, open the possibility of strengthening these approaches and make them more accurate, by exploiting vast amounts of data comprising not only events but also associated data attributes. The downside of this amount of available data is that accuracy needs to be traded off against efficiency. Thus, to fully exploit the potential of "big data" for predictive business process monitoring, powerful techniques are required to be able to (i) efficiently predict the future of several pending execution cases per time, while

(ii) still guaranteeing suitable accuracy.

In previous work [12], we presented an approach to predict whether or not a running case will fulfill a given predicate upon its completion, based both on: (i) the sequence of activities executed in a given case; and (ii) the values of data attributes after each execution of an activity in a case. However, this previous framework has significant runtime overhead (in the order of seconds or even minutes per prediction for a large log) as it builds classification models on-the-fly at runtime based on trace prefixes of completed cases that are similar to the trace of the running case. Such a technique, while achieving a good level of accuracy, is not applicable in settings with high throughput or when instantaneous response times are required to help users make rapid decisions.

In order to significantly reduce runtime overhead, the framework herein proposed builds predictive models offline, and directly applies these pre-built models at runtime. The offline component follows a two-phase approach. First, prefixes of traces of completed cases are clustered according their control flow characteristics. Secondly, for each such cluster, a classifier is constructed taking into account data attributes associated to the events in the trace prefixes. The classifier is targeted at discriminating between trace prefixes that lead to fulfillments of a predicate and those that lead to violations. The online phase consists in taking the uncompleted, evolving trace of a running case, matching it to a cluster, and applying the corresponding classifier to estimate the probability of fulfillment of the predicate at hand. The input predicates can be temporal logic constraints, or constraints on execution times, or any boolean function on completed traces.

The approach has been implemented in the ProM process mining toolset. ProM provides a generic Operational Support (OS) environment [19, 21] that allows the tool to interact with external workflow management systems at runtime. A stream of events coming from a workflow management system is received by an OS service. The OS service is connected to a set of OS providers implementing different types of analysis that can be performed online on the stream. Our *Predictive Monitoring Framework* has been implemented as an OS provider.

The paper is structured as follows. Section 2 introduces the techniques used for predictive monitoring. Next, Section 3 introduces the proposed *Predictive Monitoring Framework*. In Section 4, we show how the framework can be implemented thus producing different variants. Section 5 describes experimental results in a real-life scenario. Finally, Section 6 discusses related work and Section 7 concludes and spells out directions for future work.

## 2 Background

In this section, we provide a brief overview about the main state-of-the-art techniques used in the proposed framework.

### 2.1 Clustering

Clustering is a type of unsupervised learning problem in which a model has to be devised on top of unlabeled data. The main idea behind clustering is organizing a data set into groups (*clusters*), so that elements within a cluster are more similar to each other (according to a similarity or distance measure) than elements belonging to different clusters. A multitude of clustering algorithms have been proposed in the literature, as well as a number of possible dimensions for their classification. The *Model-based clustering* and the *Density-based spatial clustering of applications with noise (DBSCAN) clustering* are two of the most known and used of these algorithms.

Model-based clustering [8] can be seen as a generalization of the k-means clustering algorithm. It assumes that the data is generated based on a model and tries to recover it. The algorithm takes as input the number of clusters and proceeds using a two-step optimization procedure. First, for each data point $x_i$, it finds the cluster $j$ so that the density $f(x_i|\mu_j, \Sigma_j)$ is maximal, with $\mu_j$ mean and $\Sigma_j$ covariance matrix of $j$. In the second step, the parameters $\mu_j, \Sigma_j$ are recomputed based on the data points belonging to $j$.

The DBSCAN algorithm [5] belongs to the family of the density-based clustering algorithms. The idea is clustering together points having a distance between each other that is below a given threshold (i.e., with a given density), while leaving out isolated points (i.e., points that can be considered as noise). The algorithm takes as input the minimum radius of a cluster ($\epsilon$) and the minimum number of points in a cluster ($minPoints$). It starts with an arbitrary starting point and, if its $\epsilon$-neighborhood (points at a distance that is below $\epsilon$) does not contain at least $minPoints$ points, the point is temporarily marked as noise, otherwise a cluster is created. Each point $q$ in the $\epsilon$-neighborhood of the initial point is added to the cluster and, if $q$ is also dense, its $\epsilon$-neighborhood is in turn added to the cluster. The procedure iterates until the density-connected cluster is completely found. At this point a new unvisited point is analyzed, leading to the discovery of a new cluster or noise.

## 2.2 Classification

The objective of classification is learning how to assign data to predefined class labels and, more specifically, in learning a classification function $f_c$. Different classifiers (and related variants) have been proposed in the literature. Decision trees approaches (e.g., decision trees and random forests) are typical classifiers.

Decision tree learning [16] uses a decision tree as a model to predict the value of a target variable based on input variables (features). Decision trees are built from a set of training data. Each internal node of the tree is labeled with an input feature. Arcs stemming from a node labeled with a feature are labeled with possible values or value ranges of the feature. Each leaf of the decision tree is labeled with a class, i.e., a value of the target variable given the values of the input variables represented by the path from the root to the leaf. Each leaf of the decision tree is associated with a support (*class support*) and a probability distribution (*class probability*). *Class support* represents the number of examples in the training set, that follow the path from the root to the leaf and that are correctly classified; *class probability* is the percentage of examples correctly classified with respect to all the examples following that specific path.

Random forest was first introduced by Breiman [2]. At the core of the method there is still the concept of decision tree. However, instead of training a single tree on a dataset, it grows a pre-defined number of trees and let them vote for the most popular outcome. The main idea behind the algorithm is to use random selection of features for each of the trees. This way, the correlation between trees is reduced, which results in a classifier that is more robust to outliers and noise. Since each tree is trained independently, the training procedure is fast even for large data sets with many features and can be easily parallelized. Despite being considered as a "black-box" type of the method, it was put a lot of effort in the machine learning community lately to gain proper interpretation of the results [11, 14].

## 3 Predictive Monitoring

The idea of predictive monitoring is to determine whether a current running trace will comply or not to a given predicate, based on historical knowledge. As a consequence, like most of the traditional process monitoring techniques, also our framework requires a compliance model with respect to which the behavior of running traces is verified. This compliance model can include any predicate that can be evaluated over a completed trace. Therefore, for the application of the framework a classification function $f_c$, input of the framework, is needed to classify historical traces into compliant and non-compliant traces. For illustrative purposes, in our experimentation, we will use LTL on finite traces to define predicates and we take their translation into finite state automata [9] as classification functions.

In this section, first, we describe an approach for predictive monitoring presented in [12], we use as baseline for evaluating our framework for clustering-based predictive monitoring. Then, we introduce the framework itself. Based on the information extracted from past executions of the process, both techniques try to predict how currently evolving executions will develop in the future. For example, a doctor would like to predict whether "the patient will recover within one year from the diagnosis" (input predicate). To make predictions, both techniques take into consideration how this predicate was evaluated over historical traces similar to the currently running trace. This similarity is evaluated based on characteristics related to both control flow and data. We assume that data used in the process, expressed as pairs *attribute:value*, is globally visible throughout the whole process execution. We call *data snapshot* of a process activity $A$, the set of values assigned to this global set of attributes after the execution of $A$ (and of all the activities before $A$). For example, in Figure 1, the data snapshot associated to the doctor diagnosis activity ($D$) will contain not only values associated to attributes of the diagnosis activity, e.g., the diagnosis ($dia$) and possibly the prescription ($pre$) of the doctor, but also those associated to past activities, e.g., patient's symptoms ($sym$). We indicate with $DS(D) = \{sym : painA, dia : d1, pre : p1, ...\}$ the data snapshot of activity $D$.

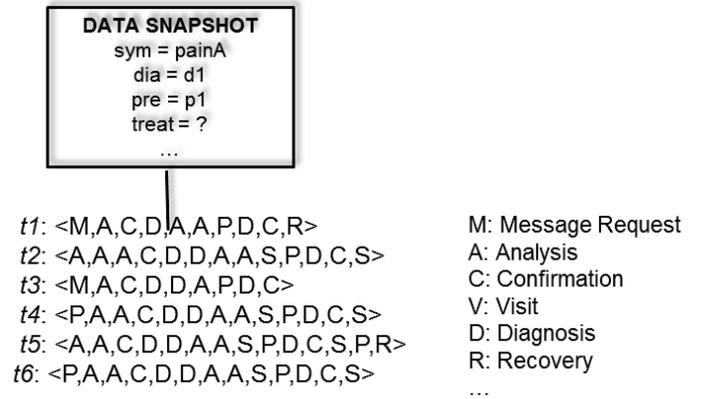

**Figure 1:** *Recovery Diagnosis Example.*

## 3.1 On-the-fly Predictive Monitoring

In this section, we present the details of the approach presented in [12]. This approach builds classification models on-the-fly at runtime based on trace prefixes of completed cases to provide predictions about the fulfillments of a predicate in an execution trace. In the following, we provide an overview of the approach and of its implementation.

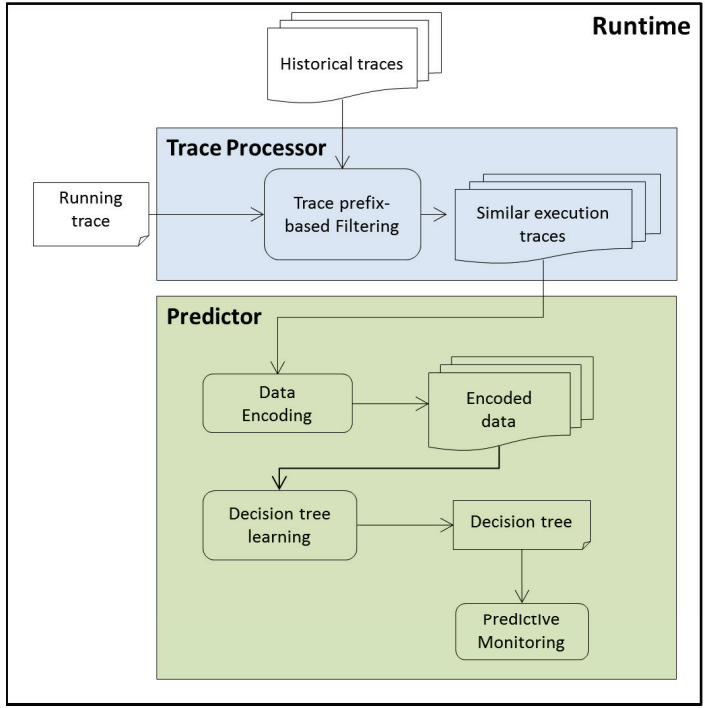

**Figure 2:** *On-the-fly predictive monitoring.*

Figure 2 sketches the approach. It relies on two main modules: a *Trace Processor* module to filter (past) execution traces and a *Predictor* module, which uses the information contained in the *Trace Processor* output as training data to provide predictions. Both modules operate at runtime.

The *Trace prefix-based Filtering* submodule of the *Trace Processor* module extracts from the set of historical traces only those traces having a prefix control flow similar to the one of the current execution trace (up to the current event). The filtering is needed since traces with similar prefixes are more likely to have, eventually in the future, a similar behavior. The similarity between two traces is evaluated based on their edit distance. We use this abstraction (instead of considering traces with a prefix that perfectly matches the current partial trace) to guarantee a sufficient number of examples to be used for the decision tree learning. In particular, a *similarity threshold* can be specified to include more traces in the training set (by considering also the ones that are less similar to the current trace).

The traces of the training set (and the corresponding selected prefixes) are then passed to the *Data Encoding* submodule, in charge of their preparation for training the decision tree. Specifically, the submodule (i) classifies each trace based on whether the desired predicate is satisfied or not (this is done by using the input classification function $f_c$); (ii) identifies for each trace the *data snapshot* containing the assignment of values for each attribute in the corresponding selected prefix. The encoding of the trace is then obtained by combining the value of the classification function on the specific trace and the trace data snapshot. For example, given the trace $t1$ in Figure 1, and its prefix $\langle M, A, C, D \rangle$, the data snapshot of the last activity in the prefix is taken. In particular, assuming that we have the vector of data attributes $\langle sym, dia, pre, treat \rangle$, the encoding for the specific trace will be the vector $\langle painA, d1, p1, ?, yes \rangle$, where the question mark is used to identify not available data values, while the last value represents the value of the classification function $f_c$ on the specific case. In this example, it represents the fact that the patient in the specific historical trace $t1$ recovered within a year from the diagnosis.

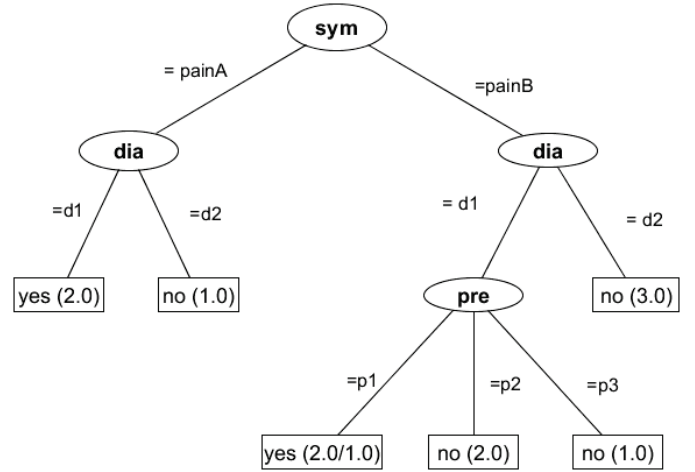

**Figure 3:** *Example decision tree.*

Once the relevant traces and, therefore, the corresponding data snapshots, are classified and encoded, they are passed to the *Decision tree learning* module, in charge of deriving the learned decision tree. The decision tree is queried using the data snapshot of the current execution trace to derive a prediction. Figure 3 shows a decision tree related to our running example. The non terminal nodes of the tree contain the decision points for the prediction (the data attributes in our case), while the arcs are labeled with possible values assigned to the attribute in the node. The leaves, instead, represent the value of the classification function on the specific path tree. The number of data training examples (with values of the input variables following the path from the root to each leaf) respectively correctly and non-correctly classified is reported on the corresponding leaf of the tree. For example, if the data snapshot of the current execution trace corresponds to values $painB$, $d1$ and $p1$, the resulting class is the formula satisfaction ("yes"), with 2 examples of the training set following the same path correctly classified (class support) and 1 non-correctly classified, i.e., with a class probability $prob = \frac{2}{2+1} = 0.66$. Therefore, in this case, the *Predictor* will predict the satisfaction of the formula with a class probability $prob = 0.66$. Note that if a path from the root to a leaf of the tree cannot be identified starting from the data snapshot of the current execution trace (e.g., if some data is missing) no prediction can be returned.

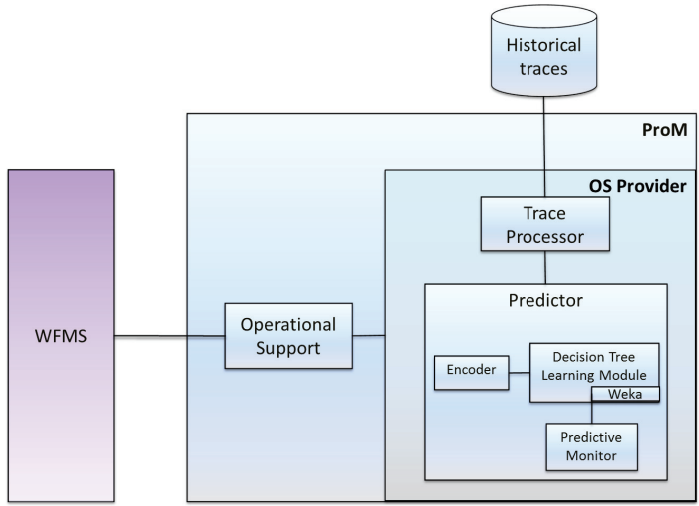

**Figure 4:** *On-the-fly predictive monitoring: implemented architecture.*

The approach has been implemented as an OS provider in ProM. Figure 4 shows the entire architecture. The OS service of ProM receives a stream of events (including the current execution trace) from a workflow management system and forwards it to the provider that returns back predictions. The OS service sends these results back to the workflow management system.

For the implementation of the *Predictor*, we rely on the WeKa J48 implementation of the C4.5 algorithm, which takes as input a .arff file and builds a decision tree. The .arff file contains a list of typed variables and, for each trace prefix (i.e., for each data snapshot), the corresponding values. This file is created by the *Encoder* and passed to the *Decision Tree Learning Module*. The resulting decision tree is then analyzed to generate predictions (*Predictive Monitor*).

## 3.2 Clustering-Based Predictive Monitoring

Differently from the approach described in the previous section, in the proposed framework, the on-the-fly construction of the decision tree can be avoided by applying a simple pre-processing phase. In such a phase, state-of-the-art approaches for clustering and classification are applied to the historical data in order to (i) identify and group historical trace prefixes with a similar control flow (clustering from a control flow perspective); and (ii) get a precise classification in terms of data of traces with similar control flow (data-based classification). At runtime, the classification of the historical trace prefixes is used to classify new traces during their execution and predict how they will behave in the future. The overall picture of the framework is illustrated in Figure 5. In the following, we describe each of the framework components in detail.

**Trace selection and encoding** Before applying state-of-the-art techniques for clustering and classification, two propaedeutical steps are applied: (i) the selection of the historical trace prefixes to consider; and (ii) their encoding. In particular, prefixes of past execution traces are selected (rather than the entire trace or all the prefixes for a trace). The reason behind this choice is twofold: on the one side, taking all the prefixes could become very expensive in terms of efficiency. On the other side, we are interested in early predictions, when still reparative actions can be undertaken to prevent violations. In this light, considering only the initial parts of the historical traces seems to be a reasonable choice. For example, given the 6 traces $t1, \ldots, t6$ of Figure 1, only a selection of $k$ prefixes for each trace will be considered. Different approaches can be used for the the selection of these $k$ prefixes. For example, the first $k$ prefixes of each historical trace can be selected or alternatively $k$ prefixes, one every $g$ events. In the latter case two prefixes differ one from another for a *gap* of $g$ events. $g$ and $k$ are user-specified parameters. Different approaches can also be taken to perform the encoding of trace prefixes for clustering. Just to name a few, a trace (prefix) can be encoded as a sequence of events or in terms of the frequency of the occurrence of sequence patterns in the trace. The simplest case is the one related to the occurrence of unary patterns, i.e., patterns composed of a single log event. For example, in the scenario in Figure 1, we can represent the alphabet of the events as an ordered vector $L = \langle A, C, D, M, P, R, S, V \rangle$. In this case, trace $t1$ will be encoded as a vector of frequencies $\langle 3, 2, 2, 1, 1, 1, 0, 0, 0 \rangle$, obtained by replacing each symbol of the alphabet in vector $L$ by its frequency in trace $t1$. Trace prefixes encoded in this way are used as input of the clustering phase. On the other hand, trace prefixes in each cluster are also used as input for supervised learning. In this case, the data perspective is taken into consideration. Historical execution traces are described using information about data, i.e., prefixes clustered based on control flow are now analyzed from a data perspective. Similarly to what explained for the on-the-fly approach, each prefix is identified with a vector that includes elements corresponding to the data assignments contained in the data snapshot associated to the last activity of the prefix. In addition, each prefix in a cluster is classified based on whether the corresponding completed trace is compliant with the input predicate (this is done by using the input classification function $f_c$).

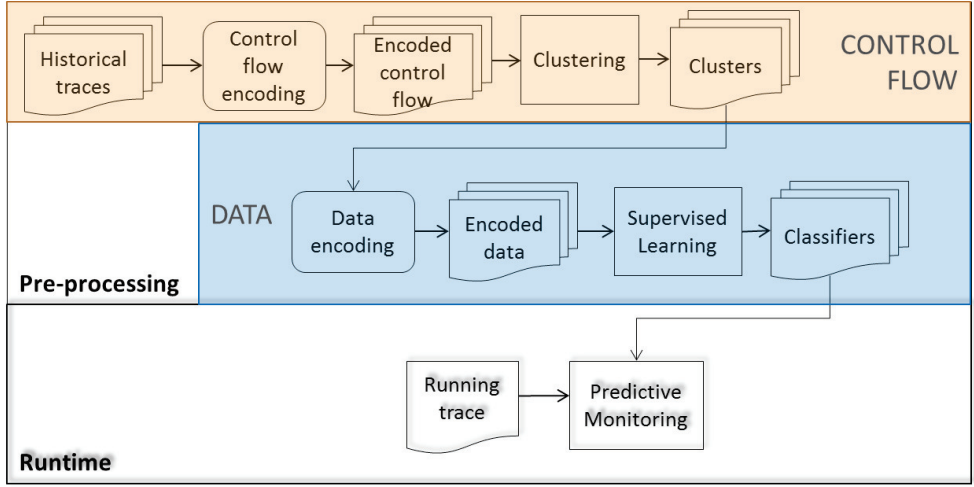

**Figure 5:** *Predictive Monitoring Framework.*

**Clustering** In the clustering phase, a selection of prefixes of the historical traces with the same (control flow) characteristics is grouped together based on some distance notion. The historical traces contained in each cluster are then used to generate a classifier, that is exploited, in turn, to make predictions on running traces, once identified their membership cluster. For example, the execution traces in Figure 1 could be grouped by a clustering algorithm in two clusters $c1$ and $c2$, according to the similarities in their control flow, so that $c1$ contains traces $t1$ and $t3$ (which have a very similar control flow), and $c2$ contains the remaining four traces.

**Supervised learning** Once traces with a similar control flow have been grouped together, their data snapshots can be used for classification. In particular, each cluster is used as training set of a supervised learning technique (e.g., decision tree learning, random forests) to generate a classifier that allows for discriminating between compliant and non-compliant behaviors. For example, given the two clusters $c1$ and $c2$, for each of them a classifier will be built.

**Predictive monitoring** At runtime, the set of classifiers generated during the pre-processing phase is used to make predictions about how the behavior of a current running trace will develop in the future. At any point in time, the current prefix of the running trace is classified as part of one of the clusters identified during the pre-processing phase. Based on the selected cluster (and, therefore, based on the control flow characteristics of the current prefix) the corresponding classifier is selected. This classifier is queried using the data snapshot of the last activity of the current prefix (exploiting the data perspective of the current prefix). For example, given a partial execution trace $tp : \langle M, A, C, D \rangle$ and the predicate "the patient will recover within a year from the diagnosis", we first identify the cluster to which the partial trace belongs, e.g., $c1$, and then the classifier associated to the cluster (e.g., the decision tree in Figure 3) is exploited in order to predict whether the predicate will be verified or not.

# 4 Implementation

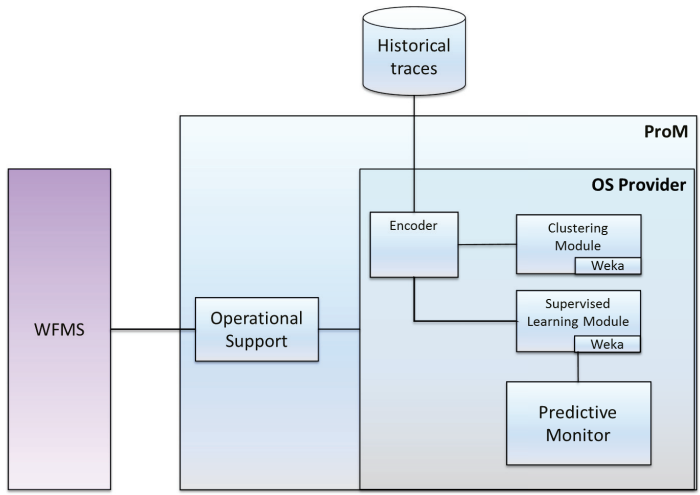

**Figure 7:** *Predictive Monitoring Framework: implemented architecture.*

The modules of the framework have been implemented by using different techniques for experimentation purposes. The clustering module has been implemented by using two different types of trace encoding and different types of clustering algorithms. In particular, a *frequency based* and a *sequence based* trace encoding approaches have been implemented. The former is realized encoding each execution trace as a vector of event occurrences (on the alphabet of the events), while, in the latter, the trace is encoded as a sequence of events. These encodings can then be passed to the clustering techniques (e.g., the ones described in Section 2). For instance, in our experiments, the *frequency based* encoding has been used with the *Model-based clustering* and the *sequence based* encoding with the *DBSCAN clustering*. In addition, for model-based clustering, we use the Euclidean distance to identify the clusters while, for DBSCAN, we use the edit distance. Finally, the supervised learning module has been implemented by using decision tree and random forest learning. The possible "instances" of our framework can be obtained through different combinations of these techniques. The implementations used in our experimentation are reported in Figure 6.

The implementations have been plugged in ProM. In particular, our *Predictive Monitoring Framework* has been implemented as an OS provider. Figure 7 shows the overall architecture.

For the implementation of the *Predictive Monitoring Framework*, we rely on (i) the Weka clustering approach implementations for some of the clustering methods, and on (ii) the WeKa J48 implementation of the C4.5 algorithm and of the random forests for the supervised learning. The result of these modules are then passed to the *Predictive Monitor*, which makes predictions on the incoming stream of events.

# 5 Evaluation

We have conducted a set of experiments by using the BPI challenge 2011 [1] event log. This log pertains to a healthcare process and, in particular, contains the executions of a process related to the treatment of patients diagnosed with cancer in a large Dutch academic hospital. The whole event log contains $1,143$ cases and $150,291$ events distributed across 623 activities. Each case refers to the treatment of a different patient. The event log contains domain specific attributes in addition to the standard XES[1] attributes such as *Age*, *Diagnosis*, and *Treatment code*.

## 5.1 Experiment Settings

The methodology adopted for the experimentation is illustrated in Figure 8. First, we have ordered the traces in the log based on the time at which the first event of each trace has occurred. Then, we split the log temporally into two parts. We used the first part (80% of the traces) as training log, i.e., we used these traces as historical data to construct clusters and build the classification models for prediction. Then we implemented a log replayer to simulate the execution of the remaining 20% of traces (the testing log) by pushing them as an event stream to the OS service in ProM and making predictions during replay. We defined 4 predicates to be used as compliance models for prediction corresponding to the following LTL rules:

---
[1]XES (eXtensible Event Stream) is an XML-based standard for event logs proposed by the IEEE Task Force on Process Mining (www.xes-standard.org).

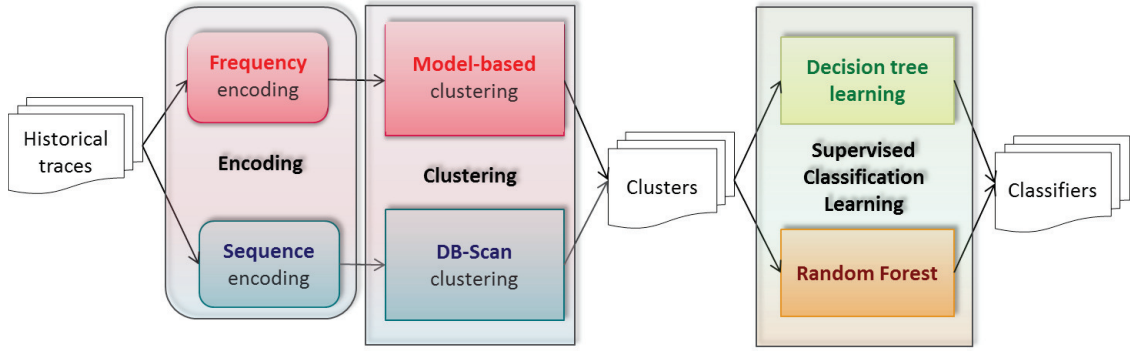

**Figure 6:** *Instances of our Predictive Monitoring Framework.*

- $\varphi_1 = \mathbf{F}(\text{``tumor marker } CA - 19.9\text{''}) \vee \mathbf{F}(\text{``}ca - 125 \text{ using meia''})$,

- $\varphi_2 = \mathbf{G}(\text{``}CEA - \text{tumor marker using meia''} \rightarrow \mathbf{F}(\text{``squamous cell carcinoma using eia''}))$,

- $\varphi_3 = (\neg\text{``histological examination } - \text{ biopsies } nno\text{''}) \mathbf{U}(\text{``squamous cell carcinoma using eia''})$.

- $\varphi_4 = \mathbf{F}(\text{``histological examination } - \text{ big resectiep''})$, and

This set of rules, indeed, allows us to exercise all the LTL constructs while investigating possibly real business constraints.

For our experiments, first, we have identified a selection of prefixes of the historical traces. In particular, the first prefixes of each historical trace have been selected, one every $g \in \{3, 5, 10\}$ events (starting from the prefix of length 1 up to the prefix of length 21). For example, in case of $g = 5$, the selected prefixes have lengths 1, 6, 11, 16, and 21 for each trace.

As described in Section 4, we use two different approaches to encode the selected prefixes: frequency based encoding and sequence based encoding. The encoded prefixes are then used for clustering. In our experiments, we have used the clustering methods described in Section 2. In particular, the model-based clustering has been performed on frequency-encoded prefixes and DBSCAN on sequence-encoded prefixes. For each of the three data sets ($g \in \{3, 5, 10\}$), we have identified the optimal parameters for clustering. In case of model-based clustering, we have chosen the optimal number of clusters using the Bayesian Information Criteria (BIC). In particular, we have applied model-based clustering with 15 to 35 clusters and chosen the value that achieved the highest BIC. The DBSCAN parameters have been estimated by using the *sorted k-dist graph* [5]. In particular, we have fixed $minPoints = 4$ and $\epsilon = 0.125$.

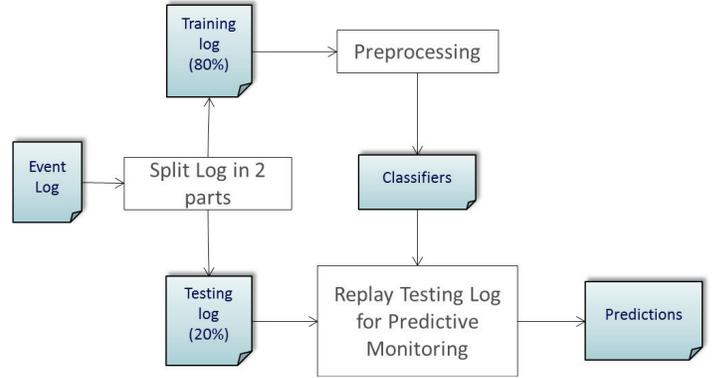

**Figure 8:** *Experimentation Methodology.*

For each cluster, we build a classifier. In particular, each cluster is used as training set of a supervised learning technique to generate a classifier that allows for discriminating between compliant and non-compliant behaviors. In our experiments, we have used decision trees and random forests.

Combining the two investigated clustering techniques and the two classification techniques, we obtain the following four *Predictive Monitoring Framework* instances:

- *mbased_dt*: it uses the model-based clustering as clustering technique and the decision tree as classification technique;

- *dbscan_dt*: it uses the DBSCAN clustering as clustering technique and the decision tree as classification technique;

- *mbased_rf*: it uses the model-based clustering as clustering technique and the random forest as classification technique;

- *dbscan_rf*: it uses the DBSCAN clustering as clustering technique and the random forest as classification technique.

We have replayed each trace in the testing set giving a prediction every 5 events (starting from the first event in each trace). Each prefix of each running trace is encoded in the same way as the historical traces and assigned to the closest cluster. In case of model-based clustering, the closest cluster is the cluster with the minimum Euclidean distance from the current prefix, while, for DBSCAN, the closest cluster is the cluster containing the prefix with the minimum edit distance from the current prefix. We use the classifier associated to the closest cluster to predict the label for the running trace. To consider a prediction reliable, the corresponding class support and class probability need to be above a given minimum class support and minimum class probability threshold. In our experiments, the minimum class support is set to $s = 6$. The minimum class probability thresholds considered are $prob \in \{0.6, 0.7, 0.8, 0.9\}$. The trace in the testing set is replayed until either a satisfactory prediction is achieved or the end of the trace is reached. In the latter case, the final prediction is considered uncertain.

## 5.2 Research Questions and Metrics

The goal of the evaluation is focused on two aspects: the performance of the approach in terms of the quality of the results and the performance of the approach in terms of the time required to provide predictions.

In particular, we are interested in answering the following two research questions:

**RQ1** Is the *Predictive Monitoring Framework effective* in providing *accurate* results as *early* as possible?

**RQ2** Is the *Predictive Monitoring Framework efficient* in providing results?

We evaluated the performance of the approach by using the following measures:

1. *accuracy*,
2. *earliness*,
3. *failure-rate*,
4. *computation time*.

In particular, *accuracy*, *earliness* and *failure-rate* have been used for answering **RQ1** and *computation time* for **RQ2**.

**Accuracy** This measure is defined with respect to a *gold standard* that indicates the correct labeling of each trace. In our experiments, we extracted the gold standard by evaluating the input predicate on each completed trace in the testing set. Given the gold standard, we classify predictions made at runtime into four categories: *i)* true-positive ($T_P$: positive outcomes correctly predicted); *ii)* false-positive ($F_P$: negative outcomes predicted as positive); *iii)* true-negative ($T_N$: negative outcomes correctly predicted); *iv)* false-negative ($F_N$: positive outcomes predicted as negative). Accuracy, which intuitively represents the proportion of correctly classified results (both positive and negative), is defined as:

$$accuracy = \frac{T_P + T_N}{T_P + F_P + T_N + F_N} \quad (1)$$

**Earliness** As already mentioned, during the replay of each trace in the testing log, we give a prediction every 5 events (starting from the first event in the trace). While replaying a trace, we consider a prediction reliable and we stop the replay when the corresponding class support and class probability are above the given minimum class support and minimum class probability thresholds. The earliness of the prediction is evaluated using the ratio between the index indicating the position of the last evaluation point (the one corresponding to the reliable prediction) and the size of the trace under examination. Notice that earliness is a crucial metrics since during the execution of a business process the stakeholders must be provided with predictions as soon as possible to apply possible reparative actions in case there is high probability of non-compliance in the future.

**Failure-rate** Sometimes it happens that, when replaying a trace of the testing set, the end of the trace is reached and no prediction has been made. In this case, the answer of the predictor is "maybe" to indicate that it was not possible to provide a reliable prediction. The percentage of traces in the log that leads to a failure in the prediction is called failure-rate.

**Computation time** We estimate three different types of computation times required for providing a prediction:

- *init time*: the time required for pre-processing, i.e., for clustering and supervised learning;
- *processing time*: the total time required for processing the entire testing set;

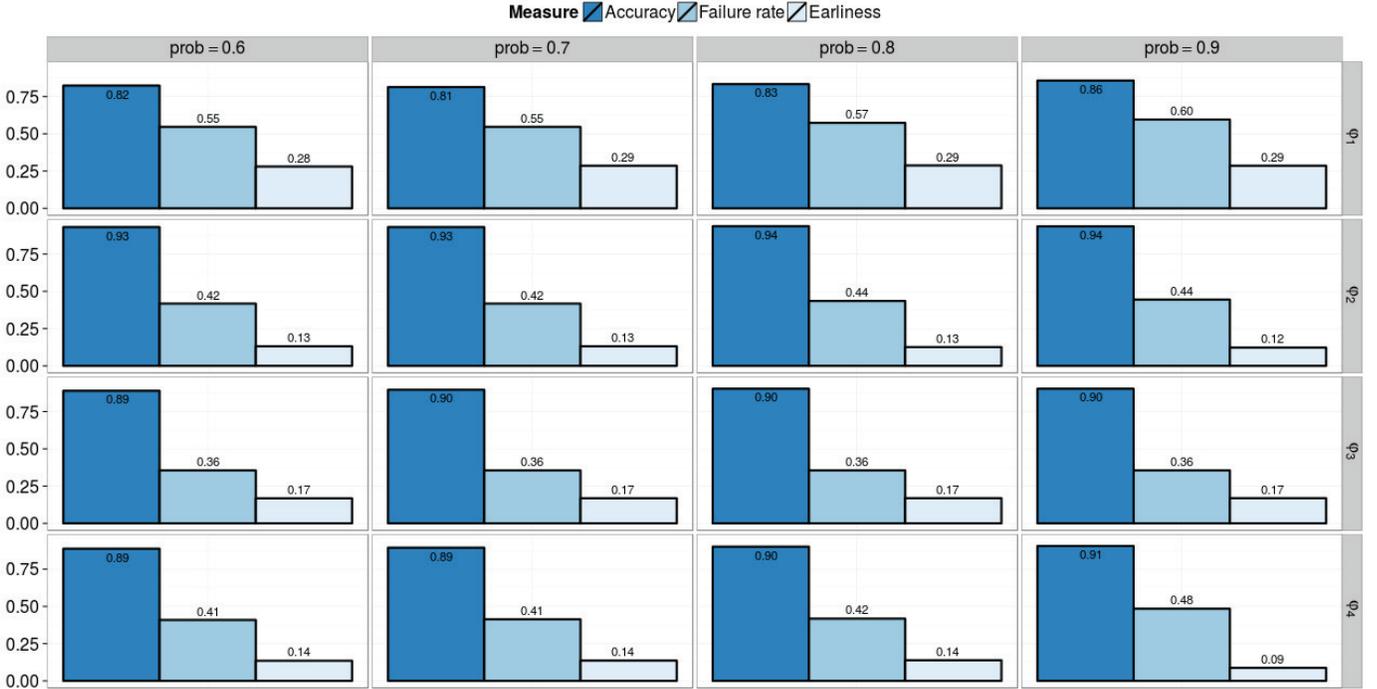

**Figure 9:** *On-the-fly predictive monitoring - accuracy, earliness and failure rate.*

- *average prediction time*: the average time required to the predictor for returning an answer at each evaluation point.

## 5.3 Results

Figure 9 reports *accuracy*, *failure rate* and *earliness* obtained by applying the baseline on-the-fly approach for each of the four investigated predicates ($\varphi 1 - \varphi 4$) with different minimum class probability thresholds. By looking at the plots, the three metrics seem not to be particularly affected by the differences in terms of minimum class probability thresholds. Overall, for the four predicates, the accuracy is reasonably high, ranging between the minimum value of $\varphi 1$ (0.81) and the maximum one of $\varphi 2$ (0.94). The failure rate is also quite high, varying from a value of 0.36 for $\varphi 3$ up to 0.6 for $\varphi 1$, while the earliness is reasonably low for $\varphi 2$, $\varphi 3$ and $\varphi 4$ and slightly higher for $\varphi 1$.

Figure 10 plots, for each of the four predicates, *accuracy*, *earliness* and *failure rate* obtained by instantiating the *Predictive Monitoring Framework* with the model-based clustering and with the decision tree classification (*mbased_dt*) for different minimum class probability values and prefix gaps. The plots show that *mbased_dt* reaches peaks of accuracy of 0.98 with a high threshold for minimum class probability ($minConf = 0.9$), though at the expenses of the failure rate (0.64). Focusing on the only results with a reasonably low failure rate (e.g., with failure rate lower than 0.25), *mbased_dt* still guarantees to find, for each predicate, a parameter configuration resulting in a good accuracy. The best accuracy values with failure rate lower than 0.25 range, indeed, between 0.85 and 0.93 for the four predicates.

Intuitively, a high minimum class probability threshold should result in a high accuracy since only reliable predictions are provided by the *Predictive Monitoring Framework*. However, this is not always the case because a too high minimum class probability can bring to discard also correct predictions (e.g., as in the case of $minConf = 0.9$ for $\varphi 3$ and $\varphi 4$).

In general, by opportunely selecting the minimum class probability threshold, it is possible to meet different needs and preferences (e.g., few but reliable predictions, many predictions although including possibly not accurate results). For instance, for *mbased_dt*, it seems that $minConf = 0.9$ ensures high values of accuracy, while $minConf = 0.8$, with the opportune choice of the prefix gap, offers a good trade-off between accuracy and failure rate for all the predicates.

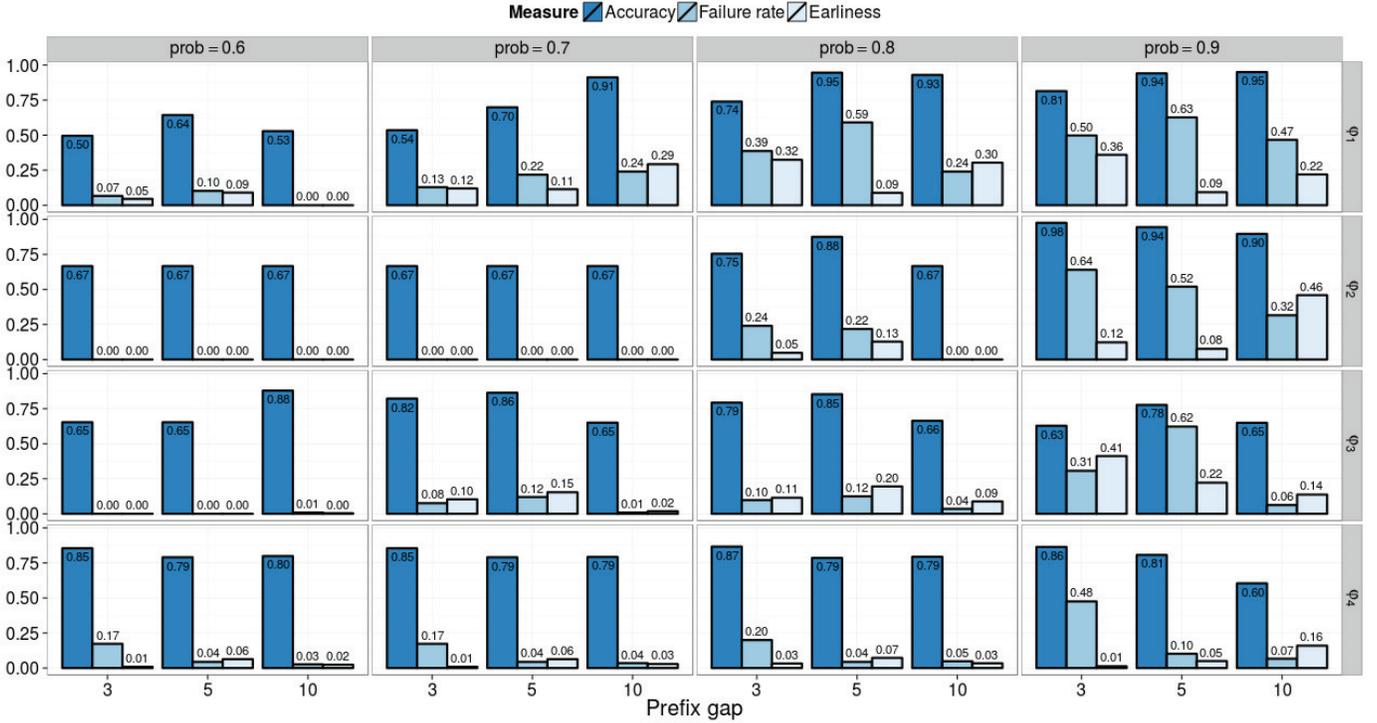

**Figure 10:** *Model-based clustering and decision tree classification (*mbased_dt*) - accuracy, failure rate and earliness.*

In most of the cases, as expected, the higher the minimum class probability threshold is, the higher failure rate and earliness are. Indeed, when the minimum class probability threshold is set to a high value, a prediction is given only if it has a high (above the minimum threshold) class probability. This means that either the trace is replayed until the observed events are enough to provide a reliable prediction (thus increasing the earliness), or that the end of the current trace is reached and no prediction is made (thus increasing the failure rate).

Figure 11 shows the same type of plots of Figure 10 obtained by instantiating the *Predictive Monitoring Framework* with DBSCAN clustering and with the decision tree classification (*dbscan_dt*). Also in this case, the approach is able to provide very high accuracy values (0.96) to the detriment of a high failure rate (0.56). In this case, looking at the only results with a low failure rate (i.e., failure rate lower than 0.25), for three out of the four predicates the best accuracy value (over the different configurations) range between 0.82 and 0.91, while it is lower (around 0.68) for the first predicate.

It seems that, for *dbscan_dt* the $minConf$ value that better balances accuracy and failure rate is 0.7.

Figure 12 shows the values of accuracy, failure rate and earliness for the model based clustering and the random forest classification (*mbased_rf*). For $minConf = 0.9$, the accuracy reaches 0.97 for two of the four predicates, while for $\varphi 3$ it is not able to perform better than 0.79 in terms of accuracy. As in the other *Predictive Monitoring Framework* instances, also in this case, ($\varphi 1$ and $\varphi 4$), high values of accuracy are accompanied by very high values of failure rate ($\sim 0.8$). By looking at the only results with failure rate lower than 0.25, also in this case it is possible to find, for each predicate, a minimum class probability threshold and prefix gap such that the accuracy values for all predicates lie in the range $0.75 - 0.79$.

Figure 13 reports the values of the three metrics for the last instance of the *Predictive Monitoring Framework*, i.e., the one using DBSCAN clustering and random forest classification (*dbscan_rf*). By observing the plots, it is possible to notice that, in this case, the accuracy values do not differ too much one from the other by varying the configuration settings: the accuracy ranges from a minimum value of 0.66 to a maximum value of 0.81. The same behavior can be observed for the earliness (for all the considered configurations it is always lower than 0.04), but it does not hold for the failure rate: the failure rate increases as the minimum class probability increases. Comparing the *Predictive Monitoring Framework* and the baseline results, it comes out that the accuracy values (and corresponding failure rates) reached with the random forest instances of the *Predictive Monitoring Framework* are comparable or higher than those obtained with the on-the-fly approach. Moreover, all the four instances, differently from the baseline, also offer the possibility to choose solutions balancing accuracy and failure rate. Summing up, from the considerations and the plots above, we can conclude that, overall, all the four considered *Predictive Monitoring Framework* instances are able to provide solutions presenting very high accuracy values (accompanied by high failure rate values), but also solutions offering a good trade-off between accuracy and failure rate already at the early events of the current trace for all the considered predicates (**RQ1**).

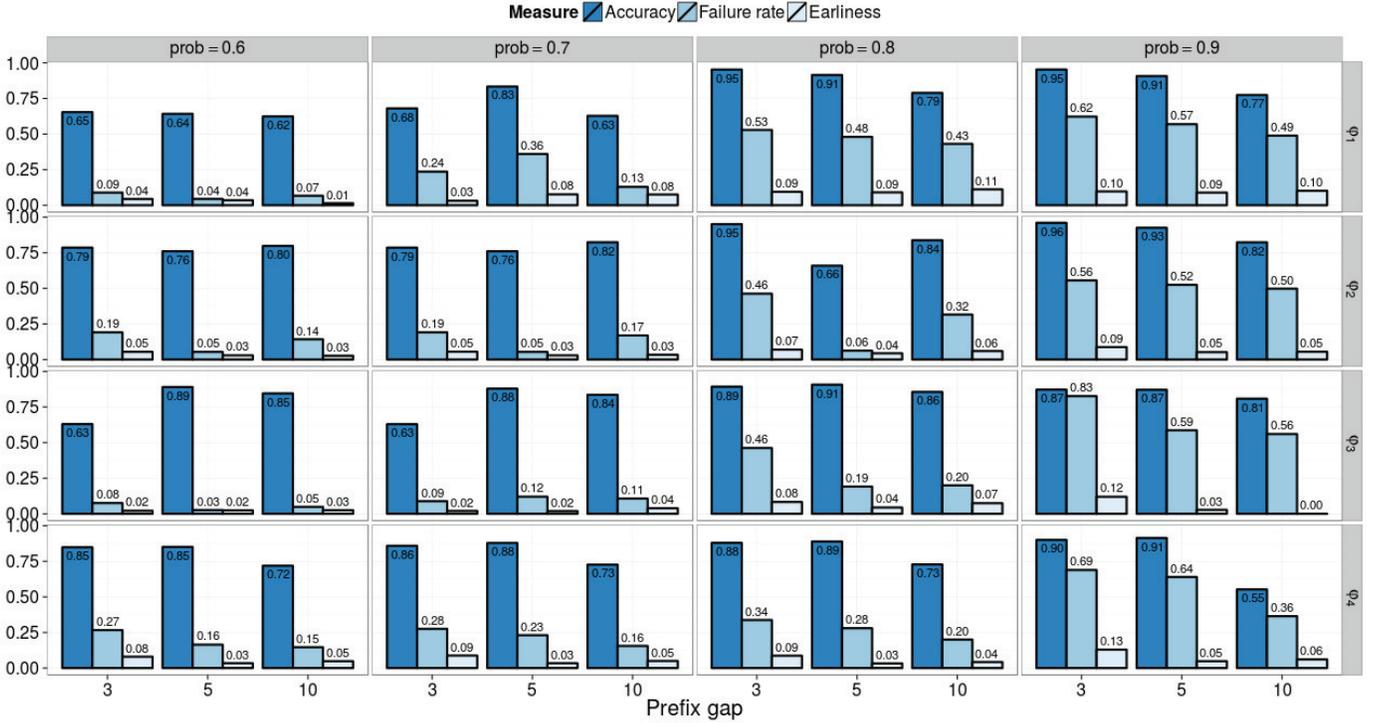

**Figure 11:** *DBSCAN clustering and decision tree classification (*dbscan_dt*) - accuracy, earliness and failure rate.*

| $\varphi$ | minConf=0.6 | | | minConf=0.7 | | | minConf=0.8 | | | minConf=0.9 | | |
|---|---|---|---|---|---|---|---|---|---|---|---|---|
| | init | processing | avg. pred. | init | processing | avg. pred. | init | processing | avg. pred. | init | processing | avg. pred. |
| $\varphi_1$ | | 453946.354 | 165.84 | | 456517.477 | 164.116 | | 466850.858 | 171.528 | | 474879.207 | 177.843 |
| $\varphi_2$ | | 368796.069 | 25.532 | | 366606.233 | 25.36 | | 368895.547 | 26.622 | | 371021.759 | 27.94 |
| $\varphi_3$ | | 395761.404 | 45.773 | | 396113.806 | 45.835 | | 395810.652 | 45.797 | | 395810.652 | 45.797 |
| $\varphi_4$ | | 234582.674 | 38.239 | | 238284.95 | 40.224 | | 247649.381 | 46.305 | | 290813.698 | 25.854 |

**Table 1:** *On-the-fly predictive monitoring - Processing and average prediction time (in seconds)*

To answer **RQ2**, we focus on two of the four *Predictive Monitoring Framework* instances, the ones based on decision tree classification, since the computational time required by the corresponding random forest ones, do not present significant differences.

Table 1 reports *init time* (which is null for the on-the-fly approach), the *processing time* and the *average prediction time* obtained by applying the on-the-fly approach for each of the four investigated predicates with different minimum class probability thresholds. The results in the table show that
12

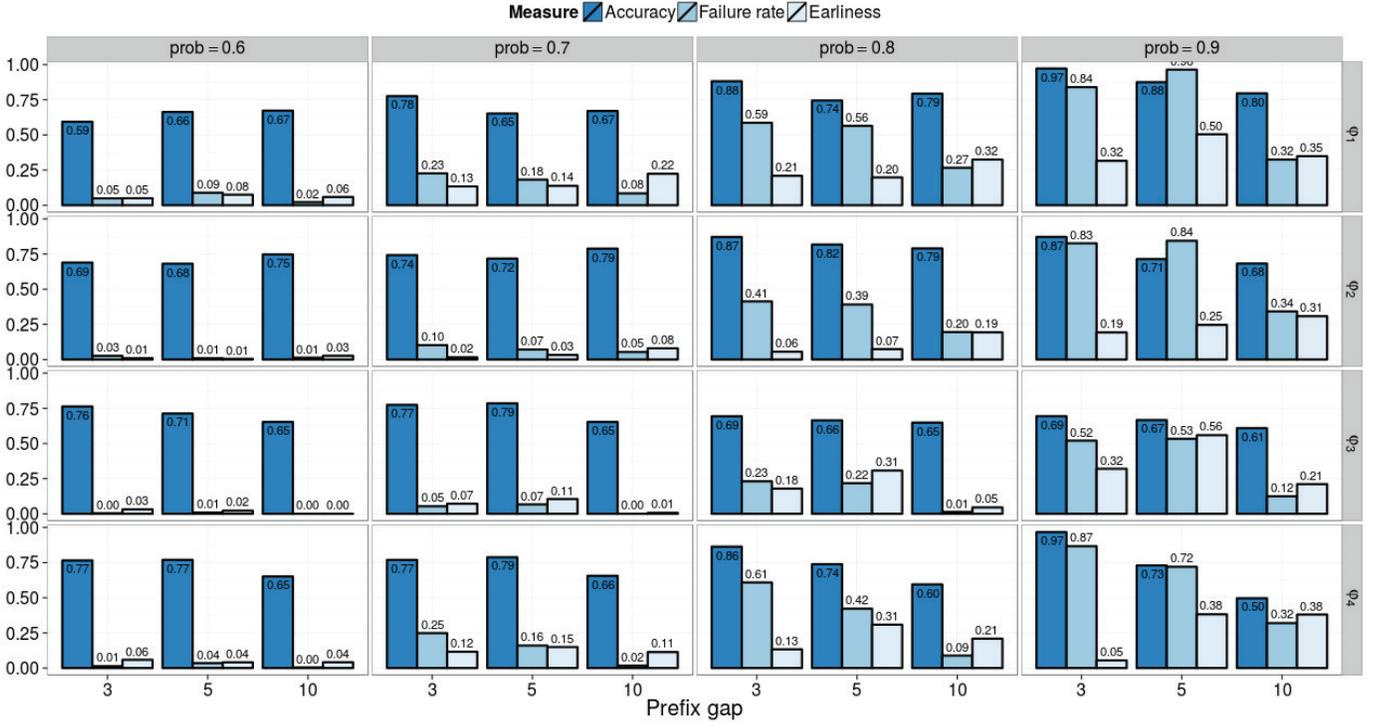

**Figure 12:** *Model-based clustering and random forest classification (*mbased_rf*) - accuracy, earliness and failure rate.*

processing a trace with such an approach is very expensive in terms of time required (the processing time ranges between $\sim 100$ to $\sim 130$ hours), which makes the approach difficult to be used in run-time scenarios. The average prediction time, i.e., the time required at each evaluation point to provide a prediction, strongly depends on the investigated predicate, and ranges from about 25 seconds for $\varphi 2$ to about 3 minutes for $\varphi 1$.

Table 2 reports the computational time, and specifically the *init time*, *processing time* and *average prediction time*, required for providing a prediction for each of the four predicates with different minimum class probability thresholds and considering different prefix gaps for *mbased_dt*. By observing the table, it seems that no big differences exist in terms of pre-processing time (*init-time*) for different predicates and minimum class probability thresholds, while it depends on the considered trace prefix gap: the higher the prefix gap is, less prefix traces are in the training set, less time is required to process them. In general, very small differences can be observed in terms of average prediction time (always $\sim 10$ milliseconds), while bigger differences occur in terms of the time required for processing a trace, ranging from a minimum value of 1 second to a maximum value of about 67 seconds. These values are, however, still reasonable to be considered for providing predictions at runtime. The differences in terms of processing time depend on the predicate, on the prefix gap as well as on the minimum class probability threshold. More in general, the processing time is related to the failure rate and the earliness: the higher the failure rate (and the earliness) are for the specific settings, the higher the number of evaluation points that have to be processed is (and the processing also increases).

Finally, Table 3 reports the *init time*, the *processing time* and the *average prediction time* required to the *Predictive Monitoring Framework* instance obtained combining the DBSCAN clustering and the decision tree classification techniques (*dbscan_dt*) for providing predictions. Also in this case, the initialization is constant for different minimum class probability thresholds and predicates, while it depends on the prefix gap (ranging from about 1.25 minutes to about 5 minutes). Small differences exist in terms of average prediction times (ranging from 8 to 31 milliseconds), while more significant ones hold in terms of processing time. The processing time, indeed, ranges between 1 to 30 minutes, when the failure rate is very high ($\varphi 3$, $minConf = 0.9$, $gap = 3$).

| $\varphi$ | minConf=0.6 ||||||||| 
| | gap=3 ||| gap=5 ||| gap=10 |||
| | init | processing | avg. prediction | init | processing | avg. prediction | init | processing | avg. prediction |
|---|---|---|---|---|---|---|---|---|---|
| $\varphi_1$ | 2844.829 | 2.54 | 0.007 | 1660.859 | 5.824 | 0.01 | 713.558 | 3.429 | 0.009 |
| $\varphi_2$ | 2844.444 | 1.366 | 0.006 | 1662.347 | 7.536 | 0.01 | 714.918 | 5.615 | 0.009 |
| $\varphi_3$ | 2843.995 | 1.378 | 0.006 | 1661.454 | 11.959 | 0.01 | 713.166 | 2.697 | 0.008 |
| $\varphi_4$ | 2844.803 | 10.95 | 0.006 | 1661.599 | 18.301 | 0.01 | 713.522 | 4.644 | 0.009 |
| $\varphi$ | minConf=0.7 |||||||||
| | gap=3 ||| gap=5 ||| gap=10 |||
| | init | processing | avg. prediction | init | processing | avg. prediction | init | processing | avg. prediction |
| $\varphi_1$ | 2842.966 | 8.138 | 0.007 | 1661.364 | 34.657 | 0.01 | 714.605 | 15.411 | 0.009 |
| $\varphi_2$ | 2843.775 | 1.344 | 0.006 | 1660.577 | 9.807 | 0.011 | 713.023 | 7.059 | 0.009 |
| $\varphi_3$ | 2843.078 | 4.861 | 0.007 | 1661.214 | 20.782 | 0.011 | 713.106 | 47.958 | 0.009 |
| $\varphi_4$ | 2845.012 | 10.938 | 0.007 | 1661.828 | 39.13 | 0.01 | 713.915 | 13.48 | 0.009 |
| $\varphi$ | minConf=0.8 |||||||||
| | gap=3 ||| gap=5 ||| gap=10 |||
| | init | processing | avg. prediction | init | processing | avg. prediction | init | processing | avg. prediction |
| $\varphi_1$ | 2843.349 | 22.655 | 0.007 | 1662.209 | 15.328 | 0.01 | 714.142 | 10.584 | 0.009 |
| $\varphi_2$ | 2843.716 | 16.625 | 0.006 | 1661.085 | 26.023 | 0.011 | 713.067 | 12.912 | 0.009 |
| $\varphi_3$ | 2845.504 | 6.267 | 0.007 | 1660.463 | 8.854 | 0.011 | 713.077 | 4.618 | 0.009 |
| $\varphi_4$ | 2844.309 | 13.942 | 0.007 | 1661.118 | 44.609 | 0.01 | 713.349 | 16.78 | 0.01 |
| $\varphi$ | minConf=0.7 |||||||||
| | gap=3 ||| gap=5 ||| gap=10 |||
| | init | processing | avg. prediction | init | processing | avg. prediction | init | processing | avg. prediction |
| $\varphi_1$ | 2844.202 | 25.645 | 0.007 | 1660.841 | 67.494 | 0.01 | 713.522 | 17.333 | 0.009 |
| $\varphi_2$ | 2844.342 | 37.222 | 0.006 | 1661.668 | 15.94 | 0.011 | 714.25 | 6.388 | 0.009 |
| $\varphi_3$ | 2844.619 | 17.646 | 0.007 | 1661.371 | 4.877 | 0.01 | 714.319 | 3.412 | 0.009 |
| $\varphi_4$ | 2846.026 | 29.548 | 0.007 | 1660.566 | 9.377 | 0.01 | 713.438 | 7.065 | 0.009 |

**Table 2:** *Model-based clustering and decision tree classification (*mbased_dt*)- Init, processing and average prediction time (in seconds)*

| $\varphi$ | minConf=0.6 |||||||||
| | gap=3 ||| gap=5 ||| gap=10 |||
| | init | processing | avg. prediction | init | processing | avg. prediction | init | processing | avg. prediction |
|---|---|---|---|---|---|---|---|---|---|
| $\varphi_1$ | 296.707 | 77.535 | 0.03 | 174.26 | 43.927 | 0.023 | 94.956 | 30.457 | 0.01 |
| $\varphi_2$ | 296.474 | 470.605 | 0.024 | 174.093 | 65.692 | 0.018 | 76.567 | 83.382 | 0.01 |
| $\varphi_3$ | 299.63 | 61.476 | 0.02 | 172.858 | 32.376 | 0.019 | 82.567 | 20.599 | 0.011 |
| $\varphi_4$ | 301.507 | 349.717 | 0.025 | 172.904 | 260.02 | 0.02 | 76.932 | 103.178 | 0.011 |
| $\varphi$ | minConf=0.7 |||||||||
| | gap=3 ||| gap=5 ||| gap=10 |||
| | init | processing | avg. prediction | init | processing | avg. prediction | init | processing | avg. prediction |
| $\varphi_1$ | 297.478 | 720.306 | 0.026 | 172.587 | 816.293 | 0.025 | 78.03 | 61.288 | 0.013 |
| $\varphi_2$ | 299.725 | 473.14 | 0.024 | 173.853 | 66.565 | 0.0187 | 75.541 | 149.526 | 0.011 |
| $\varphi_3$ | 298.888 | 74.922 | 0.021 | 173.233 | 479.454 | 0.018 | 76.189 | 107.485 | 0.012 |
| $\varphi_4$ | 297.841 | 350.111 | 0.027 | 174.746 | 603.756 | 0.019 | 78.221 | 102.639 | 0.012 |
| $\varphi$ | minConf=0.8 |||||||||
| | gap=3 ||| gap=5 ||| gap=10 |||
| | init | processing | avg. prediction | init | processing | avg. prediction | init | processing | avg. prediction |
| $\varphi_1$ | 296.225 | 1313.315 | 0.031 | 175.039 | 937.048 | 0.025 | 75.784 | 308.192 | 0.015 |
| $\varphi_2$ | 298.736 | 1567.935 | 0.026 | 173.587 | 70.065 | 0.019 | 79.096 | 591.503 | 0.014 |
| $\varphi_3$ | 294.316 | 1110.0883 | 0.027 | 174.532 | 545.689 | 0.022 | 77.114 | 176.959 | 0.015 |
| $\varphi_4$ | 301.411 | 784.481 | 0.028 | 175.386 | 642.789 | 0.019 | 75.418 | 128.4 | 0.011 |
| $\varphi$ | minConf=0.9 |||||||||
| | gap=3 ||| gap=5 ||| gap=10 |||
| | init | processing | avg. prediction | init | processing | avg. prediction | init | processing | avg. prediction |
| $\varphi_1$ | 296.36 | 1395.725 | 0.029 | 174.147 | 1047.62 | 0.025 | 75.997 | 355.925 | 0.015 |
| $\varphi_2$ | 299.012 | 1577.679 | 0.028 | 173.56 | 1294.726 | 0.023 | 78.354 | 619.202 | 0.012 |
| $\varphi_3$ | 297.687 | 2057.379 | 0.027 | 173.599 | 1018.924 | 0.019 | 76.989 | 420.525 | 0.008 |
| $\varphi_4$ | 294.808 | 1300.652 | 0.027 | 176.473 | 1041.599 | 0.018 | 78.387 | 213.356 | 0.013 |

**Table 3:** *DBSCAN clustering and decision tree (*dbscan_dt*) - Init, processing and average prediction time (in seconds)*

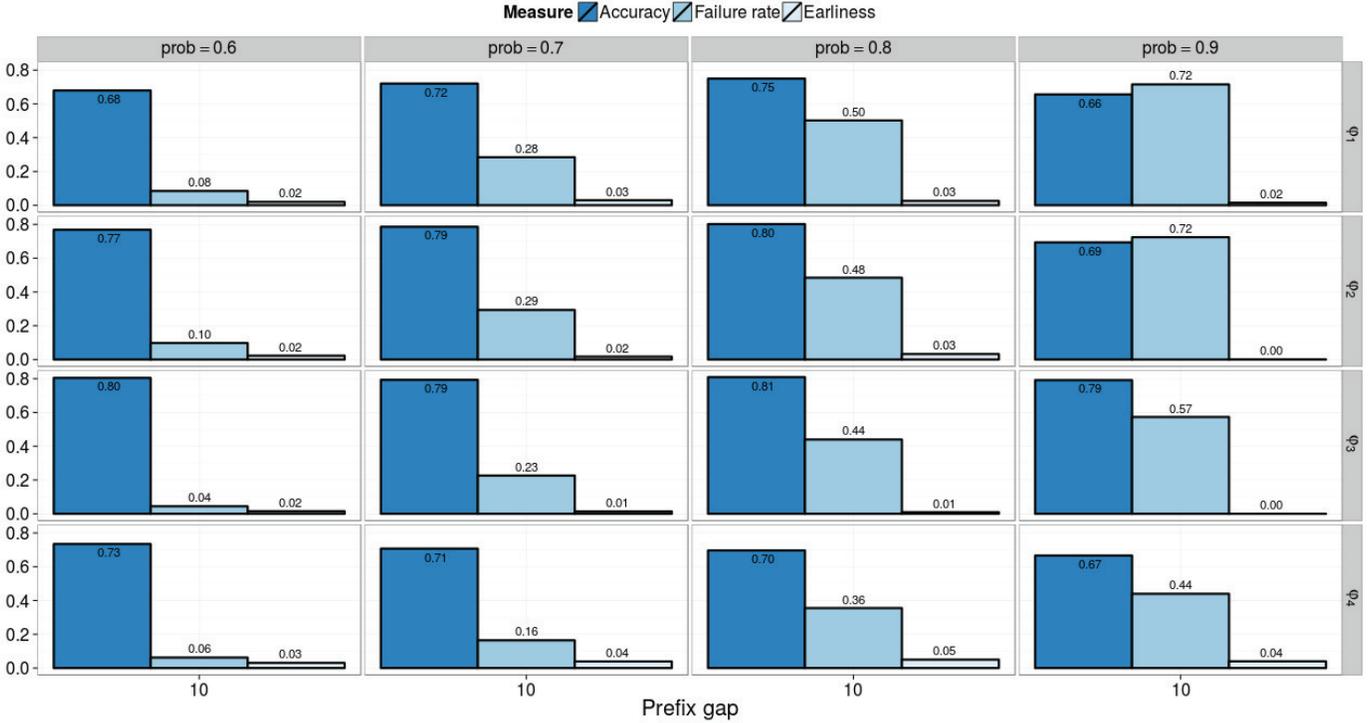

**Figure 13:** *DBSCAN clustering and random forest classification (*dbscan_rf*) - accuracy, earliness and failure rate.*

By comparing the performance of the proposed *Predictive Monitoring Framework* instances with the on-the-fly approach, it clearly comes out that, with comparable results in terms of accuracy, failure rate and earliness, the time required for processing a trace and providing a prediction is much lower with the clustering-based approach. The processing time is indeed about 300 times lower for *dbscan_dt* and 16,000 times lower for *mbased_dt*. Even taking into account the init time required by the clustering-based predictive monitoring approaches, which is computed once per all traces, such a time is anyway lower than the processing time of the on-the-fly approach. Similarly, the average prediction time for the cluster-based approaches is even 5,000 times smaller than the average prediction time of the on-the-fly approach: the average prediction time for *dbscan_dt* and *mbased_dt* is of the order of few milliseconds, while for the on-the-fly approach it can also reach two minutes. Compared with the baseline the *Predictive Monitoring Framework* provides, hence, solutions quite efficient, which can be reasonably used for providing predictions at runtime (**RQ2**).

## 5.4 Discussion

The observations and the analysis preformed so far allow us to draw some conclusions and guidelines. The solutions provided by the different instances of the *Predictive Monitoring Framework* offer the possibility to meet different types of needs, by opportunely setting the available configuration parameters.

For instance, in settings in which users are more interested in the accuracy of their results (e.g., in medical scenarios in which predictions cannot be wrong), a high minimum class probability value should be selected, while lower values should be preferred if a good trade-off between accuracy and failure rate is desired. In case, instead, users would prefer to have a prediction even if not always correct rather than a non-prediction, low minimum class probability values would allow them to get an almost null failure rate with an acceptable accuracy in many cases. The choice of the minimum configuration value however also depends on the predicate under consideration, because in some cases a too high minimum class probability threshold can cut off also correct predictions. For instance, in the investigated settings, the trend showing an increase of the accuracy in correspon-dence to high minimum class probability thresholds seems overall to apply for predicates $\varphi1$ and $\varphi2$, but not for $\varphi3$ and $\varphi4$.

Concerning the gap intervals for the prefixes used in the training, users more interested in obtaining accurate results though at the expenses of a higher computational cost, should prefer small prefix gaps, while, if they are willing to sacrifice accuracy for a faster response, bigger prefix gaps have to be selected. Also in this case, the choice of the gap interval would depend on the predicate (and on the minimum class probability threshold). For instance, in the analysed setting, for $minConf = 0.6$, the accuracy values do not vary too much for $\varphi2$, while they decrease when the prefix gap increases for $\varphi4$.

For the choice of the clustering and the classification technique to use for instantiating the *Predictive Monitoring Framework*, no big differences seems to exist in terms of peaks of accuracy, the only exception being *dbscan_rf* presenting a highest accuracy (0.81) lower than the others. However, some differences among the four instances can be identified by considering not only the accuracy but also failure rate and earliness. In particular, the instances based on random forests seem to perform lightly worse than the ones based on decision tress. Moreover, random forest instances look less sensitive to the adopted clustering technique than the decision tree ones: no significant differences can be observed in the results of *mbased_rf* and *dbscan_rf*. On the contrary, when decision trees are used as classification technique, *dbscan_dt* results seem to outperform the results obtained with *mbased_dt*.

## 6 Related Work

In the literature, there are works dealing with approaches for the generation of predictions, during process execution, focused on the time perspective. In [20, 19], the authors present a set of approaches in which annotated transition systems, containing time information extracted from event logs, are used to: (i) check time conformance while cases are being executed, (ii) predict the remaining processing time of incomplete cases, and (iii) recommend appropriate activities to end users working on these cases. In [7], an ad-hoc predictive clustering approach is presented, in which context-related execution scenarios are discovered and modeled through state-aware performance predictors. In [17], the authors use stochastic Petri nets for predicting the remaining execution time of a process.

Another group of works in the literature focuses on approaches that generate predictions and recommendations to reduce risks. For example, in [4], the authors present a technique to support process participants in making risk-informed decisions, with the aim of reducing the process risks. Risks are predicted by traversing decision trees generated from the logs of past process executions. In [15], the authors make predictions about time-related process risks, by identifying (using statistical principles) and exploiting indicators observable in event logs that highlight the possibility of transgressing deadlines. In [18], an approach for Root Cause Analysis through classification algorithms is presented. Decision trees are used to retrieve the causes of overtime faults on a log enriched with information about delays, resources and workload.

An approach for prediction of abnormal termination of business processes is presented in [10]. Here, a fault detection algorithm (local outlier factor) is used to estimate the probability of a fault to occur. Alarms are provided for early notification of probable abnormal terminations. In [3], Castellanos et al. present a business operations management platform equipped with time series forecasting functionalities. This platform allows for predictions of metric values on running process instances as well as for predictions of aggregated metric values of future instances (e.g., the number of orders that will be placed next Monday). Predictive monitoring focused on specific types of failures has also been applied to real case studies. For example, in [13, 6], a technique is presented to predict "late show" events in transportation processes by applying standard statistical techniques to find correlations between "late show" events and external variables related to weather conditions or road traffic.

A key difference between these approaches and our technique is that they rely either on the control-flow or on the data perspective for making predictions at runtime, whereas we take both perspectives into consideration. In addition, we provide a general, customizable framework for predictive process monitoring that is flexible to be implemented in different tool variants with different techniques.

## 7 Conclusion

We presented a framework for predictive monitoring of business processes that exploits data from past execution traces (both control flow and data attributes associated to events) to estimate the probability that a given predicate will be fulfilled upon completion of a running case. The framework achieves relatively low runtime overhead by constructing classification models offline – one per cluster of prefixes of historical traces.

At runtime the prediction is made by matching the running case to a cluster, and applying the corresponding classification model to extract a prediction. Compared to a previous method that computes classification models at runtime [12], this leads to comparable results in terms of accuracy and to a significant improvement in terms of response times. Experimental results show that the framework can achieve high levels of earliness (i.e., predictions are made early during the running case) and low failure rates (i.e., low number of cases where predictions cannot be made with sufficient class probability).

As the experiments are based on a single log, the results have low generalizability. Accordingly, as future work we plan to conduct further experiments across different application domains. Also, we plan to investigate the application of techniques for extraction of predictive sequence patterns [22] to enhance prediction accuracy. Yet another avenue for future work is to experiment with other clustering methods, for example hierarchical clustering [8], that could potentially lead to further improvements in terms of accuracy.

**Acknowledgments.** This research is partly funded by the ERDF via the Estonian Centre of Excellence in Computer Science.